\documentclass[a4paper,reprint,aps,prx,superscriptaddress]{revtex4-1}
\usepackage{amsmath}
\usepackage[dvipdfmx]{graphicx}
\usepackage{here}
\usepackage{bm}
\usepackage{newtxtext}
\usepackage{newtxmath}
\usepackage{mathrsfs}
\usepackage{braket}
\usepackage[colorlinks=true,allcolors=blue]{hyperref}
\usepackage[para,online,flushleft]{threeparttable}
\newcommand{\ooba}[1]{\textcolor{red}{}}
\newcommand{\tadano}[1]{#1}

\newcommand{\del}[1]{}

\graphicspath{{./figure/}}

\begin{document}
%
\title{First-principles study of phonon anharmonicity and negative thermal expansion in ScF$_{3}$}

\author{Yusuke Oba}
\thanks{Present Address: TDK Corporation, 160 Miyazawa, Minami-Alps, Yamanashi 400-0495, Japan}
\email{yusuke_oba@jp.tdk.com}
\affiliation{Department of Physics, The University of Tokyo, Hongo, Bunkyo-ku, Tokyo 113-0033, Japan}

\author{Terumasa Tadano}
\email{Tadano.Terumasa@nims.go.jp}
\affiliation{International Center for Young Scientists (ICYS), National Institute for Materials Science, Tsukuba 305-0047, Japan}
\affiliation{Research and Services Division of Materials Data and Integrated System (MaDIS), 
             National Institute for Materials Science, Tsukuba 305-0047, Japan}
            
\author{Ryosuke Akashi}
\affiliation{Department of Physics, The University of Tokyo, Hongo, Bunkyo-ku, Tokyo 113-0033, Japan}

\author{Shinji Tsuneyuki}
\affiliation{Department of Physics, The University of Tokyo, Hongo, Bunkyo-ku, Tokyo 113-0033, Japan}
\affiliation{Institute for Solid State Physics, The University of Tokyo, Kashiwa, Chiba 277-8581, Japan}

\date{\today}

\begin{abstract}
The microscopic origin of the large negative thermal expansion of cubic scandium trifluorides (ScF$_{3}$) is investigated by performing a set of anharmonic free-energy calculations based on density functional theory. We demonstrate that the conventional quasiharmonic approximation (QHA) completely breaks down for ScF$_{3}$ and the quartic anharmonicity, treated nonperturbatively by the self-consistent phonon theory, is essential to reproduce the observed transition from negative to positive thermal expansivity and the hardening of the R$^{4+}$ soft mode with heating. In addition, we show that the contribution from the cubic anharmonicity to the vibrational free energy, evaluated by the improved self-consistent phonon theory, is significant and as important as that from the quartic anharmonicity for robust understandings of the temperature dependence of the thermal expansion coefficient. The first-principles approach of this study enables us to compute various thermodynamic properties of solids in the thermodynamic limit with the effects of cubic and quartic anharmonicities. Therefore, it is expected to solve many known issues of the QHA-based predictions particularly noticeable at high temperature and in strongly anharmonic materials.
\end{abstract}
\maketitle

%
%
\section{Introduction}
\label{sec:intro}



Negative thermal expansion (NTE) materials are useful for technological applications and have been studied actively in the recent two decades~\cite{martin2016negative}. So far, various kinds of NTE materials have been discovered including metal oxides~\cite{mary1996negative, sanson2006negative, korthuis1995negative, evans1997negative}, metal fluorides~\cite{greve2010pronounced, chatterji2011negative, kennedy2002powder}, and metal-organic frameworks~\cite{hohn2002sr, birkedal2002observation}. Also, NTE materials provide a route for near-zero thermal expansion composite materials, which are desired in the field of high-precision measurements and semiconductor devices. 
The mechanism of NTE can be roughly categorized into two types: the vibrational effect and the nonvibrational effect~\cite{barrera2005negative}. The former, which applies to many NTE materials, attributes NTE to existence of phonon modes having a large negative Gr\"{u}neisen parameter~\cite{Fang:2014cp}. Indeed, many NTE materials comprise rigid octahedral or tetrahedral structure units, mainly formed by oxygen atoms, and their rotational modes often show the pressure-induced softening.

\begin{figure}[tb]
 \centering
 \includegraphics[width=5.0cm,clip]{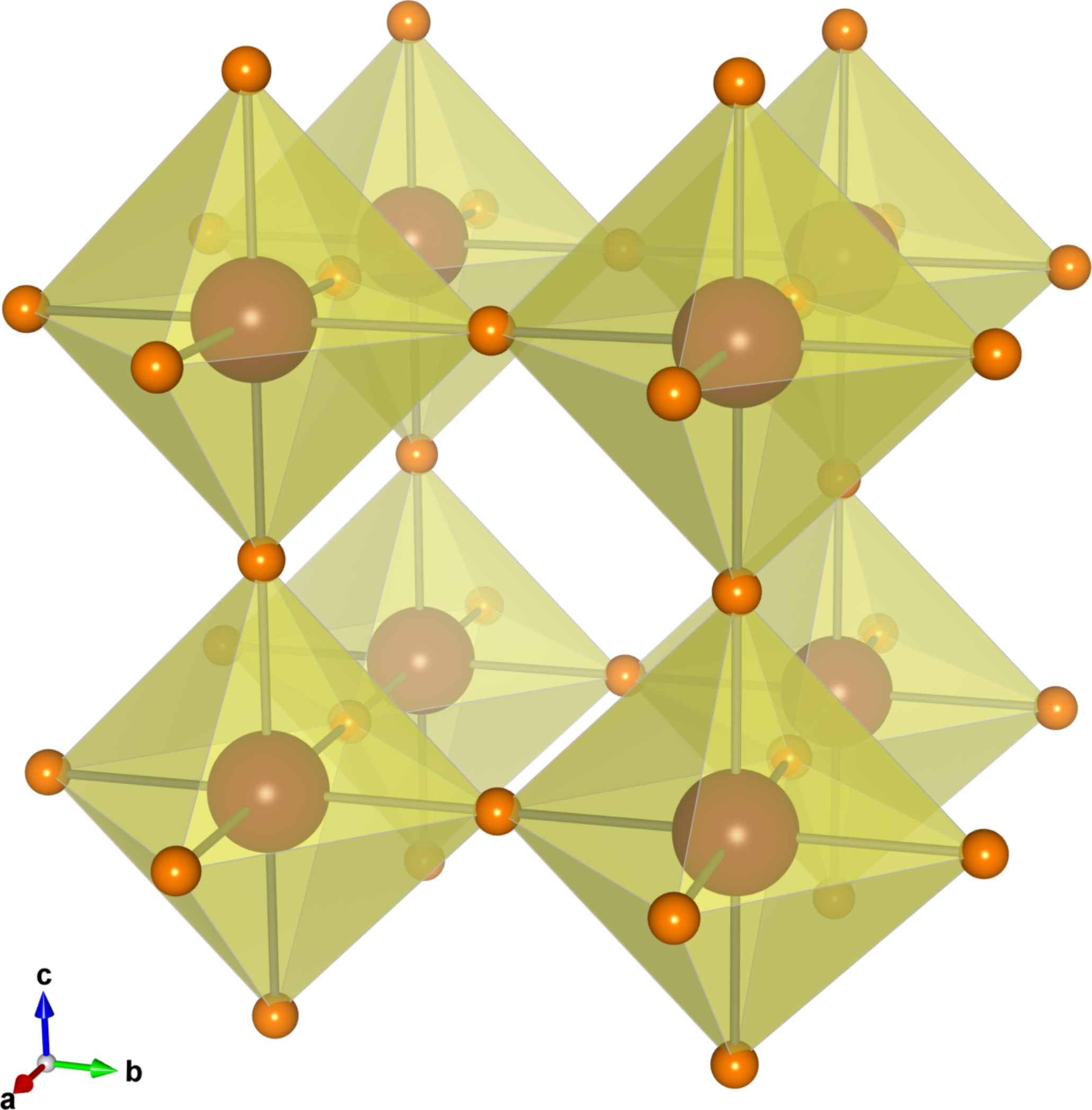}
 \caption{Crystal structure of cubic ScF$_{3}$ with space group {\it Pm$\bar{3}$m} (created with \textsc{vesta}~\cite{Momma:db5098}). Each scandium atom is surrounded by six fluorine atoms, which form corner-sharing octahedra. The structure is similar to the {\sl ABX$_{3}$} cubic perovskite, but the {\sl A} site is vacant.}
 \label{fig:ScF3}
\end{figure}

Scandium trifluorides ($\mathrm{ScF_3}$) is an NTE material belonging to the family of metal trifluorides~\cite{hepworth1957crystal, daniel1990raman}, whose structure is the $\mathrm{ReO_3}$-type ``open perovskite'' (Fig.~\ref{fig:ScF3}). Greve \textit{et al.} reported that the coefficient of the thermal expansion (CTE) of ScF$_{3}$ is strongly negative at low temperature ($\alpha_{l} \simeq -10$ ppm $\cdot$ K$^{-1}$ at 200 K), and the NTE is persistent up to $\sim$1100 K~\cite{greve2010pronounced}. Although the modest NTE can also be observed in other ReO$_{3}$-type structures~\cite{chatterji2008negative,Morelock:2014eg}, the magnitude of NTE is far lower than in ScF$_{3}$. Moreover, ScF$_{3}$ is unique in that it does not show a structural phase transition even at 0.38 K~\cite{Romao:2015fp} 
\del{\ooba{unless adding the pressure of $\sim$0.1GPa at 50K~\cite{greve2010pronounced}}},
while the other $3d$ metal trifluorides transform into a rhombohedral structure with cooling~\cite{Handunkanda:2015dc}. In ScF$_{3}$, the R$^{4+}$ soft mode, relevant to the cubic-to-rhombohedral structural phase transition, softens with cooling but does not condensate in the low-temperature limit, as evidenced by recent inelastic x-ray scattering (IXS) experiments~\cite{Handunkanda:2015dc,Occhialini:2017cv}. 
\tadano{Since cubic ScF$_{3}$ is in proximity to the rhombohedral phase at low temperatures, the cubic-to-orthorhombic phase transition can be induced by applying low pressure of $\sim$0.1 GPa at 50 K~\cite{greve2010pronounced}.}

%

To elucidate the unique thermophysical properties of ScF$_{3}$, accurate treatment of lattice dynamics including anharmonic effects is essential.
Since the discovery of the large NTE in ScF$_{3}$, several first-principles studies based on density functional theory have reported the CTE calculated within the quasi-harmonic approximation (QHA)~\cite{Li_Structural_2011,Liu:2015gj,Wang:2015gm}. Since the QHA only accounts for the volume dependence of phonon frequencies and neglects higher-order anharmonicities, however, none of these QHA results correctly reproduced the experimental CTE values. Li \textit{et al.}~\cite{Li_Structural_2011} have demonstrated that the inclusion of quartic anharmonicity of the R$^{4+}$ soft mode is significant and essential for explaining the experimental result. More recently, an improved treatment of anharmonic effects either by \textit{ab initio} molecular dynamics (AIMD)~\cite{lazar2015negative} or by a stochastic implementation~\cite{roekeghen2016anomalous} of the self-consistent phonon (SCP) theory~\cite{Werthamer_Self_1970} has been shown to explain the observed CTE more quantitatively and highlighted the importance of higher-order anharmonicities. Despite the success of these numerical methods for reproducing the temperature dependence of CTE of ScF$_{3}$ somewhat quantitatively, the microscopic mechanism of the large NTE is not fully understood. In particular, the role of the quartic anharmonicity on the CTE has not been quantified in comparison with that of the cubic anharmonicity, which can also affect the vibrational free energy. Therefore, an alternative approach is desired.

In this paper, we present another first-principles approach for incorporating anharmonic effects in the vibrational free energy beyond the QHA level. 
Our approach is based on the recent implementation of the SCP theory that uses fourth-order interatomic force constant (IFC)~\cite{Tadano_Self_2015}. 
Since the SCP approach can compute an anharmonic phonon quasiparticle nonperturbatively on a dense momentum grid, thermodynamic properties in the thermodynamic limit can be evaluated accurately, which is essential for accurate determination of phase boundaries~\cite{PhysRevLett.112.058501}. By applying the method to ScF$_{3}$, we show that the inclusion of the quartic anharmonicity solves the issue of the QHA and reproduces the experimentally observed change from negative to positive CTE with heating. In addition, we find that the effect of the cubic anharmonicity, evaluated by the improved self-consistent (ISC) theory~\cite{goldman1968improved}, is equally important and compensates the overcorrection made by the SCP theory.

This paper is organized as follows. First, we introduce the theoretical methods to incorporate anharmonic effects in vibrational free energy in Sec.~\ref{sec:theory}. We then describe computational and technical details of the DFT calculation in Sec.~\ref{sec:detail}. In Sec.~\ref{sec:QHA}, we demonstrate the limitation of the QHA theory in ScF$_{3}$, which can be resolved by more accurate treatment of anharmonic effects as shown in Sec.~\ref{sec:SCPandISC}. Moreover, we discuss the temperature dependence of the R$^{4+}$ soft mode in Sec.~\ref{sec:phonons}. Finally, a concluding remark is made in Sec.~\ref{sec:conclusion}.

%
%
\section{Vibrational free energy}
\label{sec:theory}
The Helmholtz free energy of a nonmetallic system is expressed as a function of the volume of the unit cell $V$ and temperature $T$,
\begin{equation}
F(V,T) = E_{\mathrm{el}}(V) + F_{\mathrm{vib}}(V,T),
\end{equation}
where $E_{\mathrm{el}}(V)$ is the static internal energy of electrons obtained by a first-principles calculation and
$F_{\mathrm{vib}}(V,T)$ is the vibrational free energy. 
Assuming that the lattice vibration is well described by thermal excitation of phonons,
we examine the behavior of $F_{\mathrm{vib}}(V, T)$ with the three levels of approximation in this study: the QHA, SCP theory, and ISC theory.
Once the free energy is obtained at various volumes, the $F(V,T)$ curve is fitted by the equation of state (EOS) to estimate an equilibrium volume at each temperature. By repeating the procedure at different temperatures, we can obtain the volume thermal expansion $\alpha_{\mathrm{v}}(T) = \frac{1}{V} (\frac{\partial V}{\partial T})_{P}$. For isotropic systems, the CTE is equal to $\frac{1}{3}\alpha_{\mathrm{v}}$.

For the sake of brevity, we use the shorthand notation of $q=(\bm{q},j)$ and $-q=(-\bm{q},j)$ where
$\bm{q}$ is the phonon momentum and $j$ is the phonon branch index.

\subsection{Quasiharmonic theory}
%
The quasiharmonic (QH) theory is the standard approximation of $F_{\mathrm{vib}}(V,T)$, in which the vibrational free energy is given as
\begin{equation} 
F_{\mathrm{vib}}^{(\rm{QH})}(V,T) = \frac{1}{\beta} \sum_{q} \ln \left[ 2 \sinh{\Big( \frac{1}{2}\beta\hbar\omega_{q}(V)\Big)} \right]. \label{eq:free_QHA}
\end{equation}
Here, $\beta = 1/kT$ with the Boltzmann constant $k$, and $\omega_{q}(V)$ is the phonon frequency at volume $V$ obtained within the harmonic approximation (HA).
While the anharmonic effect is partially incorporated via the volume dependence of $\omega_{q}(V)$,
the QHA completely neglects the intrinsic anharmonic effects which are responsible for making the temperature dependence of phonon frequencies. 
Nevertheless, the QH theory turns out to be a good approximation at temperatures far below the melting point and has been employed to 
predict the thermal expansivity and phase boundary of various materials based on DFT. 
When the temperature reaches the melting point or the structure is strongly anharmonic, the QHA is less reliable.
Moreover, the QHA is not valid for cases where phonon modes become unstable within the HA as will be discussed in Sec.~\ref{sec:QHA}.

%
\subsection{Self-consistent phonon theory} 
%
The SCP theory is one of the most successful approaches for calculating the temperature dependent phonon frequencies nonperturbatively~\cite{Werthamer_Self_1970, Tadano_Self_2015}.
In this method, we assume the existence of \textit{effective} harmonic phonon frequency $\Omega_{q}$ ($\Omega_{q}^{2}\geq 0$) and polarization vector $\bm{\varepsilon}_{q}$,
with which we can define an \textit{effective} harmonic Hamiltonian $\mathscr{H}_{0}$ as 
\begin{equation}
  \mathscr{H}_{0} = \frac{1}{2}\sum_{q}\hbar\Omega_{q}\mathcal{A}_{q}\mathcal{A}_{q}^{\dagger}.
\end{equation}
Here, $\mathcal{A}_{q} = a_{q} + a_{-q}^{\dagger}$ is the displacement operator with $a_{q}^{\dagger}$ and $a_{q}$ being creation and annihilation operators of SCP, respectively. 
In the first-order SCP theory, the renormalized phonon frequencies and eigenvectors $\{\Omega_{q},\bm{\varepsilon}_{q}\}$ are determined so
that the vibrational free energy within the first-order cumulant approximation is minimized. 
Let $F_{\mathrm{vib}}$ denotes the vibrational free energy of an anharmonic system described by Hamiltonian $H = H_{0} + U_{2} + U_{3} + \dots$ [see Eq.~(\ref{eq:Un})] and $F_{\mathrm{vib}}'$ denotes the free energy with the first-order cumulant approximation, the following inequality then holds:
\begin{equation}
F_{\mathrm{vib}} \leq F_{\mathrm{vib}}' = \mathscr{F}_{0} + \braket{H - \mathscr{H}_{0}}_{\mathscr{H}_{0}}.
\end{equation}
Here, $\mathscr{F}_{0} = -\frac{1}{\beta}\log{Z}$ and $\braket{X}_{\mathscr{H}_{0}} = Z^{-1}\mathrm{Tr}(Xe^{-\beta\mathscr{H}{_{0}}})$ with $Z = \mathrm{Tr}(e^{-\beta\mathscr{H}_{0}})$ the partition function and $\beta = 1/kT$.
It is straightforward to show that $F_{\mathrm{vib}}'$ is given as follows:
\begin{align}
F_{\mathrm{vib}}' &= \frac{1}{\beta} \sum_{q} \ln \left[ 2 \sinh{\Big( \frac{1}{2}\beta\hbar\Omega_{q}(V, T)\Big)} \right] \notag \\
& \hspace{5mm} + \frac{1}{2}\sum_{q} \left[ (C_{\bm{q}}^{\dagger}\Lambda_{\bm{q}}^{(\mathrm{HA})}C_{\bm{q}})_{jj} - \Omega_{q}^{2} (V,T) \right] \alpha_{q} \notag \\
& \hspace{5mm} + \frac{1}{8}\sum_{q,q'}\Phi^{(\mathrm{SCP})}(q;-q;q';-q')\alpha_{q}\alpha_{q'}, \label{eq:F1}
\end{align}
where $\Lambda^{(\mathrm{HA})}_{\bm{q}} = \mathrm{diag} (\omega_{\bm{q}1}^{2},\dots,\omega_{\bm{q}j}^{2})$ and $\alpha_{q} = \frac{\hbar}{2\Omega_{q}}[1+2n(\Omega_{q})]$ with $n(\omega)$ being the Bose--Einstein distribution function. $\Phi^{(\mathrm{SCP})}(q;-q;q';-q')$ is the fourth-order IFC in the normal coordinate basis of the SCP. For the detailed expression, see Appendix \ref{sec:TEP}. 
From the condition of $\frac{\partial F_{\mathrm{vib}}'}{\partial \Omega_{q}} = 0$ and $\frac{\partial F_{\mathrm{vib}}'}{\partial C_{q}^{\dagger}} = 0$, we obtain the first-order SCP equation as
\begin{equation}
\Omega_{q}^{2} = (C_{\bm{q}}^{\dagger}\Lambda_{\bm{q}}^{(\mathrm{HA})}C_{\bm{q}})_{jj} + \frac{1}{2}\sum_{q'} \Phi^{(\mathrm{SCP})}(q;-q;q';-q')\alpha_{q'}. \label{eq:SCP}
\end{equation}
Since the anharmonic interaction is relatively short-range compared with the harmonic one, it is sufficient to solve the above equation on an irreducible set of $\bm{q}$ points that are commensurate with an employed supercell. Once the change of the dynamical matrix $\Delta D(\bm{q}) = D^{(\mathrm{SCP})}(\bm{q})-D^{(\mathrm{HA})}(\bm{q})$ is obtained on these grids, $\Delta D(\bm{q})$ can be interpolated to arbitrary $\bm{q}$ points by the Fourier interpolation~\cite{Tadano_Self_2015}.

In Eqs.~(\ref{eq:F1}) and (\ref{eq:SCP}), we neglected the sixth and higher-order anharmonic terms assuming that their effects are far smaller than the dominant fourth-order term. With this assumption, the third term in Eq.~(\ref{eq:F1}) can be removed by using Eq.~(\ref{eq:SCP}) as 
\begin{align}
F_{\mathrm{vib}}^{(\mathrm{SCP})}(V,T) &= \frac{1}{\beta} \sum_{q} \ln \left[ 2 \sinh{\Big( \frac{1}{2}\beta\hbar\Omega_{q}(V, T)\Big)} \right] \notag \\
& - \frac{1}{4}\sum_{q} \left[ \Omega_{q}^{2} (V,T) - (C_{\bm{q}}^{\dagger}\Lambda_{\bm{q}}^{(\mathrm{HA})}C_{\bm{q}})_{jj}  \right] \alpha_{q}.  \label{eq:Free_SCP}
\end{align}
Hence evaluating the anharmonic free energy $F_{\mathrm{vib}}^{(\mathrm{SCP})}(V,T)$ in the thermodynamic limit is feasible owing to the interpolation technique, which is hardly achievable by the thermodynamic integration based on AIMD. The first term in Eq.~(\ref{eq:Free_SCP}) corresponds to the QH contribution [Eq.~(\ref{eq:free_QHA})] but the harmonic frequency is replaced with the SCP frequency. The second term is a correction necessary to satisfy the correct thermodynamic relationship $S = -dF/dT$~\cite{Allen:2015dm}, where the entropy is given as $S = k\sum_{q}[(n_{q}+1)\ln (n_{q}+1)-n_{q}\ln n_{q}]$ with $n_{q} = n(\Omega_{q})$. When the anharmonic effect is small, we can assume $\Omega_{q}\approx \omega_{q} + \Delta_{q}$ and $C_{\bm{q},ij} \approx \delta_{ij}$. Then, we obtain the small perturbation limit of Eq.~(\ref{eq:Free_SCP}) as 
\begin{align}
F_{\mathrm{vib}}^{(\mathrm{PT})}(V,T) &= \frac{1}{\beta} \sum_{q} \ln \left[ 2 \sinh{\Big( \frac{1}{2}\beta\hbar\Omega_{q}(V, T)\Big)} \right] \notag \\
& \hspace{15mm} - \frac{\hbar}{2}\sum_{q} \Delta_{q} \bigg( n_{q} + \frac{1}{2} \bigg).  \label{eq:Free_PT}
\end{align}
This result is the same as the one reported by Allen~\cite{Allen:2015dm}. In this study, we always use Eq.~(\ref{eq:Free_SCP}) because it is more generalized and even applicable to systems where unstable (imaginary) phonon modes exist within the HA.

\subsection{Improved self-consistent phonon theory}
%
The SCP theory accounts for the quartic anharmonicity nonperturbatively but neglects the effect of the cubic anharmonicity. The ISC theory~\cite{goldman1968improved} accounts for the additional three-phonon term perturbatively as 
\begin{equation}
F_{\mathrm{vib}}^{(\mathrm{ISC})}(V,T) = F_{\mathrm{vib}}^{(\mathrm{SCP})}(V,T) + F_{\mathrm{vib}}^{(\mathrm{B})}(V,T),
\label{eq:FE_ISC}
\end{equation}
where $F_{\mathrm{vib}}^{(\mathrm{B})}(V,T)$ is the Helmholtz free energy from the bubble diagram~\cite{Tadano_Self_2015}, associated with the cubic anharmonicity, given as
\begin{widetext}
\begin{align} 
F_{\mathrm{vib}}^{(\rm{B})}(V,T) 
&= -\frac{\hbar^2}{48} \sum_{q_{1},q_{2},q_{3}} \frac{\left | \Phi^{(\mathrm{SCP})} (q_{1};q_{2};q_{3}) \right |^2}{\Omega_{q_{1}}\Omega_{q_{2}}\Omega_{q_{3}}} \Delta(\bm{q}_1+\bm{q}_2+\bm{q}_3) \notag \\ 
&\hspace{10mm} \times \Biggl[\frac{(1+n_1)(1+n_2+n_3) + n_2 n_3}{\Omega_{q_{1}}+\Omega_{q_{2}}+\Omega_{q_{3}}} 
+ 3 \frac{n_1 n_2 - n_2 n_3 + n_3 n_1 + n_1}{-\Omega_{q_{1}}+\Omega_{q_{2}}+\Omega_{q_{3}}}\Biggl]. 
\label{eq:3rd}
\end{align}
\end{widetext}
Here, we simply denote \tadano{$n(\Omega_{q_{i}})$} as $n_{i}$ and $\Delta(\bm{q})=N_{q}\delta_{\bm{q},m\bm{G}}$ with any integer $m$. In the calculation of $F_{\mathrm{vib}}^{(\mathrm{B})}(V,T)$, we use the SCP lattice dynamics wavefunction instead of those within the HA. Previous numerical studies using empirical model potentials showed that the ISC theory could describe various thermodynamic properties of copper~\cite{Cowley:1974ta} and noble-gas solids~\cite{goldman1968improved,Kanney:1975wk} accurately in a wide temperature range. In this study, we evaluate all the components entering the free energy with fully nonempirical first-principles calculation.

The bubble free energy is theoretically related to the bubble self-energy $\Sigma_{q}^{(\mathrm{B})}(i\omega_{m})$ as 
\begin{equation}
F_{\mathrm{vib}}^{(\rm{B})}(V,T) = - \frac{1}{6\beta}\sum_{q}\sum_{m} G^{(\mathrm{SCP})}_{q}(i\omega_{m})\Sigma_{q}^{(\mathrm{B})}(i\omega_{m}),
\end{equation}
where $\omega_{m}=2\pi m/\beta\hbar$ is the Matsubara frequency, $G^{(\mathrm{SCP})}_{q}(i\omega_{m})= (i\omega_{m}+\Omega_{q})^{-1}-(i\omega_{m}-\Omega_{q})^{-1}$ is the SCP Green's function, and 
\begin{align} 
\label{eq:Self_2nd}
&\Sigma_{q}^{(\mathrm{B})}(i\omega_m)=
\frac{1}{16}\sum_{q_{1},q_{2}} \frac{\hbar |\Phi^{(\mathrm{SCP})} (-q; q_{1};q_{2})|^{2}}{\Omega_{q}\Omega_{q_{1}}\Omega_{q_{2}}} \notag \\
&\hspace{30mm} \times \Delta(-\bm{q}+\bm{q}_1+\bm{q}_2) f(1, 2, i\omega_m), \\
&f(1, 2, i\omega_m) =\sum_{\sigma=-1,1}\sigma \Biggl [\frac{1+n_1+n_2}{i\omega_m+\sigma(\Omega_{q_{1}} + \Omega_{q_{2}})} \notag \\
&\hspace{35mm}-\frac{n_1 - n_2}{i\omega_m+\sigma(\Omega_{q_{1}} -\Omega_{q_{2}})}\Biggl ].
\end{align}
The ISC Green's function $G^{(\mathrm{ISC})}_{q}(\omega)$ defined as 
\begin{equation}
[G^{(\mathrm{ISC})}_{q}(\omega)]^{-1} = [G^{(\mathrm{SCP})}_{q}(\omega)]^{-1} - \Sigma_{q}^{(\mathrm{B})}(\omega) \label{eq:Dyson}
\end{equation}
provides information of lattice dynamics in the same approximation level as Eq.~(\ref{eq:FE_ISC}).
Equation (\ref{eq:Dyson}) can be used to calculate the phonon spectral function with the effects of cubic and quartic anharmonicities, as demonstrated in cubic SrTiO$_{3}$~\cite{Tadano:2018ex}, and the imaginary part of $\Sigma_{q}^{(\mathrm{B})}(\Omega_{q})$ is essential for thermal conductivity calculations~\cite{Tadano_Self_2015,Tadano:2018ex,Tadano:2018is}.

Although the ISC theory is computationally more costly than the SCP theory because of the triplet $(q_{1},q_{2},q_{3})$ loop in Eq.~(\ref{eq:3rd}), its computational complexity is the same as that of thermal conductivity calculations with three-phonon interactions, which has been applied to various materials so far including relatively complex ones. Therefore, we expect the ISC is feasible even for complex systems. Moreover, for high-symmetry structures such as ScF$_{3}$, we can drastically reduce the computational cost by utilizing the symmetry of $\Phi(q; q_{1};q_{2})$~\cite{Chaput:2011dw}.

\section{Computational details}
\label{sec:detail}
%
\subsection{DFT calculation}
%
We employed \textit{Vienna ab initio simulation package} (\textsc{vasp})~\cite{kresse1996efficient} for calculating the electron states of ${\rm ScF_3}$, which implements the projector augmented wave (PAW)~\cite{Kresse1999,PAW1994} method. 
The adopted PAW potentials treat Sc $3s^{2}3p^{6}3d^{1}4s^{2}$ and F $2s^{2}2p^{4}$ as valence states~\cite{PP_note}.
A kinetic energy cutoff of 700 eV and the 12$\times$12$\times$12 Monkhorst-Pack $\bm{k}$-point mesh were employed.
The self-consistent field loop was continued until the total electronic energy change between two steps became smaller than $10^{-8}$ eV. 
To investigate the influence of the exchange-correlation functional, we adopted the local-density approximation (LDA), Perdew-Burke-Ernzerhof (PBE)~\cite{perdew1996generalized} functional, and a variant of PBE optimized for solids (PBEsol)~\cite{perdew2008restoring}.

%
\subsection{Force constant calculation} \label{subsec:IFCs}
%
To compare the vibrational free energy within the different levels of approximation, either the QH, SCP or ISC theory, it is necessary to extract second-, third-, and fourth-order IFCs from first-principles calculations. To this end, we employed the real-space supercell method with the 2$\times$2$\times$2 supercell (32 atoms), which is sufficiently large to allow the out-of-phase tilting motion of fluorine octahedra.
To extract second-order IFCs, an atom in the supercell was displaced from its equilibrium site by 0.01 \AA{} and the Hellmann--Feynman forces were calculated for the displaced configuration. From the data sets comprising the displacements and forces, we estimated the second-order terms by the least-squares fitting~\cite{Esfarjani_Method_2008}, as implemented in the \textsc{alamode}~\cite{tadano2014anharmonic,alamode} package. To check the convergence of the calculation, we changed the supercell size to 4$\times$4$\times$4 (256 atoms) and calculated the second-order terms. 
However, the calculated $F_{\mathrm{vib}}^{(\rm{QH})}(V,T)$ was almost unaltered by this change, thus validating the use of the 2$\times$2$\times$2 supercell.

The anharmonic IFCs were extracted by using the compressive sensing lattice dynamics method~\cite{Zhou_Lattice_2014}. Following the prescription given in Refs.~\onlinecite{Zhou_Lattice_2014,Tadano_Self_2015}, we generated 60 random displacement patterns from the trajectory of\del{\textit{ab initio} molecular dynamics} \tadano{AIMD} at 500 K and calculated atomic forces for these patterns. We then fitted the Taylor expansion potential (TEP)~\cite{Tadano_Self_2015} by using the least absolute shrinkage and selection operator (LASSO) with a regularization parameter optimized through the cross validation. The adopted TEP includes anharmonic terms up to the sixth order; the fourth-order IFCs are restricted to on-site, two-body, and three-body terms and the higher-order IFCs are restricted to on-site and two-body terms. We also verified the accuracy of the obtained TEP for independent test data sets. 
In the LASSO regression step, we fixed the second-order IFCs to the precalculated values and optimized the anharmonic terms only.

%
\subsection{Calculation of phonons and CTE}
%
The SCP equation [Eq.~(\ref{eq:SCP})] was solved using a numerical algorithm of Ref.~\cite{Tadano_Self_2015}, as implemented in the \textsc{alamode} code. 
The $\bm{q}$ mesh of the SCP was set to 2$\times$2$\times$2, which is commensurate with the supercell size, 
and the inner $\bm{q}'$ mesh was increased up to 8$\times$8$\times$8 to achieve convergence of anharmonic phonon frequencies. 
After the solution to the SCP equation was found, we converted the \textit{effective} dynamical matrices at the commensurate $\bm{q}$ points into the real-space \textit{effective} second-order IFCs, which was then used to calculate anharmonic phonon frequencies at a denser $\bm{q}$-point grid for Eqs.~(\ref{eq:Free_SCP}) and (\ref{eq:3rd}). For the calculation of the vibrational free energy, we employed 20$\times$20$\times$20 $\bm{q}$ point, which was sufficient to reach convergence. In all of the phonon calculations, the non-analytic correction to the dynamical matrix was included by using the Ewald's method~\cite{1997PhRvB..5510355G} with the dielectric tensor and Born effective charges obtained from density functional perturbation theory (DFPT)~\cite{baroni2001phonons} calculations. 

To calculate the CTE of ScF$_{3}$, we changed the lattice constant from the optimized value from $-0.05$ \AA{} to $+0.02$ \AA{} in steps of $0.005$ \AA{} and repeated the procedure described above and in Sec.~\ref{subsec:IFCs}. 
The lattice constants and Bulk modulus at finite temperature were estimated by fitting the obtained $F(V,T)$ curves with the Birch-Murnaghan EOS~\cite{PhysRev.71.809}, as implemented in \textsc{ase}~\cite{ase-paper}. 

%
%
\section{RESULTS and DISCUSSIONS}

\subsection{Ground state structural property}

Table~\ref{table:alat} shows the equilibrium lattice constant $a_{0}$ and bulk modulus of ScF$_{3}$ calculated from $E_{\mathrm{el}}(V)$, which are compared with previous experimental and computational results. All of the adopted functionals give $a_{0}$ values close to the experimental results of Greve \textit{et al.}~\cite{greve2010pronounced} within an error of $\sim$1\%. Among the three functionals, PBEsol best reproduces the experimental result with deviation as small as 0.1\%. In contrast, the bulk modulus estimated from $E_{\mathrm{el}}(V)$ overestimates the experimental value at 300--500 K~\cite{Morelock:2013gi}, in accord with the previous computational studies~\cite{Liu:2015gj,lazar2015negative}. 
This indicates the essential role of lattice vibration in determining $B_{0}$, which will be discussed further in Sec.~\ref{sec:SCPandISC}.

\begin{table}
\caption{Comparison of calculated and experimental lattice constant and bulk modulus of ScF$_{3}$.}
\label{table:alat}
\begin{ruledtabular}
\begin{threeparttable}
\begin{tabular}{lcccc}
   & LDA & PBE & PBEsol & Expt. \\
   \hline
 $a_{0}$ (\AA) & & & & \\
 Present & 3.986 & 4.074 & 4.031 & 4.026~\tnote{a} \\
 Ref.~\onlinecite{Liu:2015gj} & & 4.069 & & \\
 Ref.~\onlinecite{lazar2015negative} & & 4.037 & & \\
 \hline
 $B_{0}$ (GPa)& & & & \\
 Present & 106 & 89 & 96 & $\sim$ 60~\tnote{b} \\
 Ref.~\onlinecite{Liu:2015gj} & & 89 & & \\
 Ref.~\onlinecite{lazar2015negative} & & 97 & & \\
\end{tabular}
\begin{tablenotes}[flushleft]\footnotesize
\item[a] Ref.~[\onlinecite{greve2010pronounced}], $T = $ 0 K\\ 
\item[b] Ref.~[\onlinecite{Morelock:2013gi}], $T = $ 300--500 K
\end{tablenotes}
\end{threeparttable}
\end{ruledtabular}
\end{table}

\subsection{Volume dependent phonon frequency and breakdown of QHA} \label{sec:QHA} 
%

\begin{figure}[bt]
 \centering
 \includegraphics[width=7.5cm,clip]{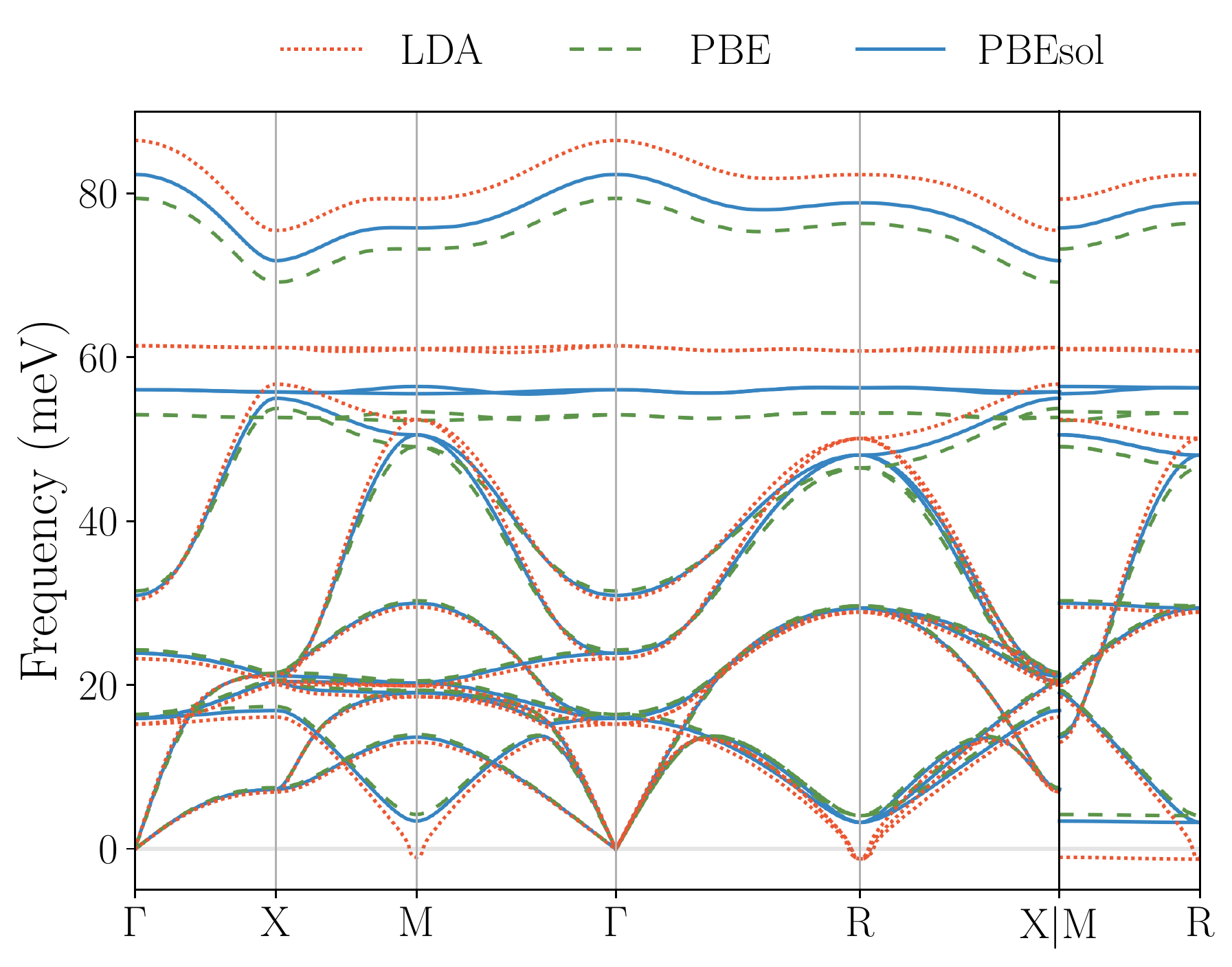}
 \caption{Comparison of harmonic phonon dispersion curves calculated with the lattice constants close to the optimized values: 3.985, 4.075, and 4.030 \AA{} for LDA, PBE, and PBEsol, respectively.}
 \label{fig:HarmonicPhonon}
\end{figure}

In this section, we demonstrate the breakdown of the QHA in ScF$_{3}$. 
Figure \ref{fig:HarmonicPhonon} compares phonon dispersion curves along high-symmetry lines calculated with LDA, PBE, and PBEsol at volumes close to their equilibrium values. In ScF$_{3}$, two soft modes exist: the triply degenerate R$^{4+}$ mode and the M$^{3+}$ mode. 
Along the high-symmetry line M-R, the soft mode shows almost no dispersion. Also, the soft mode is unstable at the equilibrium volume within LDA.

\begin{figure}[tb]
 \centering
 \includegraphics[width=6.5cm,clip]{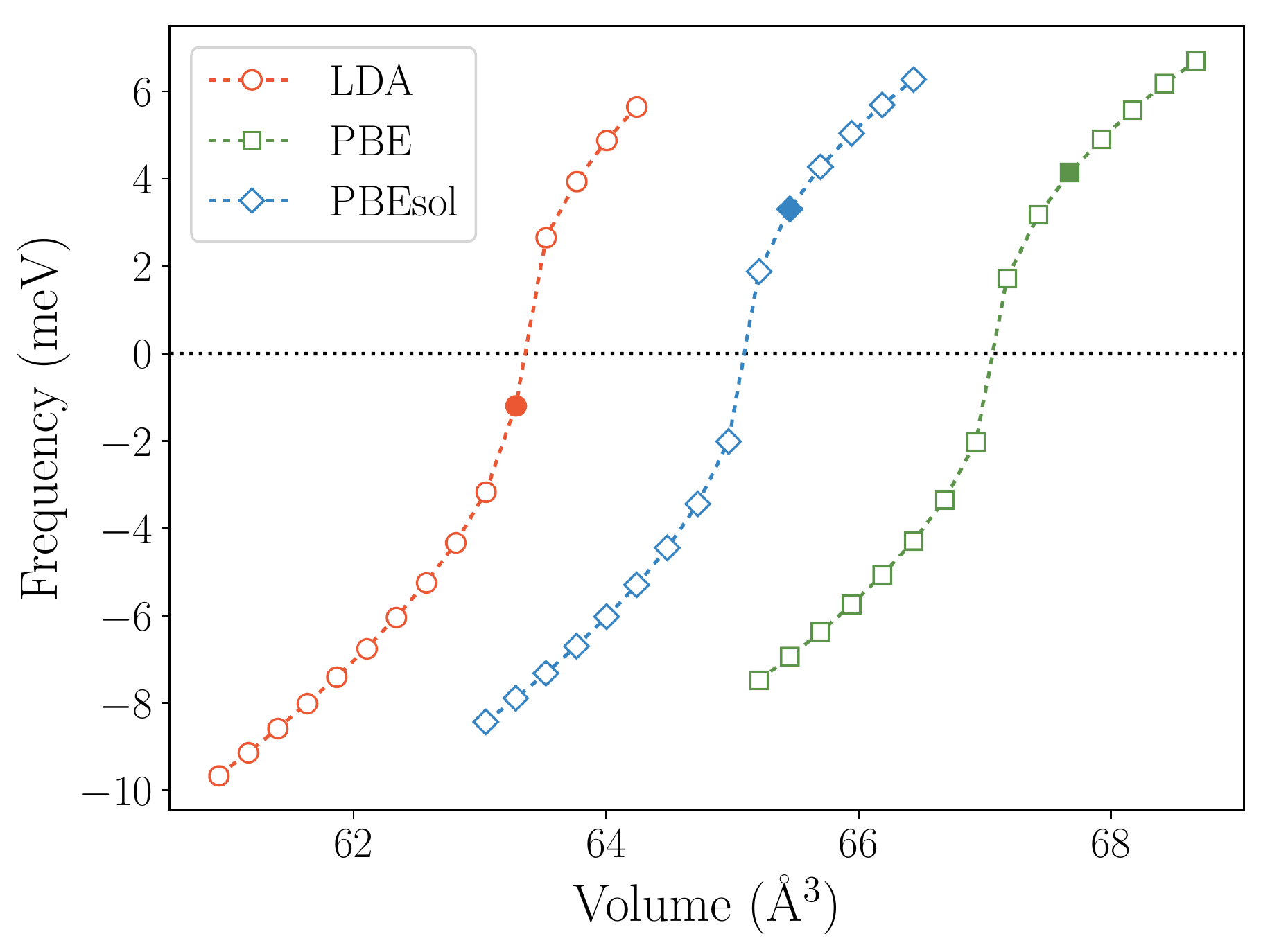}
 \caption{Volume dependence of the phonon frequency of the R$^{4+}$ soft mode. The imaginary frequencies are shown as negative values.
 The filled symbols indicate the volumes corresponding to the results shown in Fig.~\ref{fig:HarmonicPhonon}.}
 \label{fig:SoftModeVolume}
\end{figure}

The frequencies of the soft modes sensitively change with the volume as shown in Fig.~\ref{fig:SoftModeVolume}. 
The frequency of the R$^{4+}$ soft mode increase sharply with increasing the cell volume, which represents a large negative mode Gr\"{u}neisen parameter $\gamma_{q} = -\frac{V}{\omega_{q}}\frac{\partial\omega_{q}}{\partial V}$ as pointed out in the previous theoretical work~\cite{Li_Structural_2011}.
Assuming that all phonon modes are stable, the CTE is given as
\begin{align}
&\alpha_{\mathrm{v}}(T)=\frac{C_\mathrm{v} \gamma}{B_T V}, \label{eq:CTE_gamma}\\
&\gamma=\frac{\sum_{q} c_{q}\gamma_{q}}{\sum_{q} c_{q}},
\end{align}
where $C_\mathrm{v}$, $B_T$, $\gamma$, and $c_{q}$ are the isothermal heat capacity, isothermal bulk modulus, the average Gr\"uneisen parameter, and the mode-specific heat, respectively. Therefore, the NTE originates from a negative $\gamma$, which in turn can be attributed to the large negative $\gamma_{q}$ of the low-frequency soft mode whose thermal weight $c_{q}$ is relatively large in a low-temperature region. 

\begin{figure}[tb]
 \centering
 \includegraphics[width=8.5cm,clip]{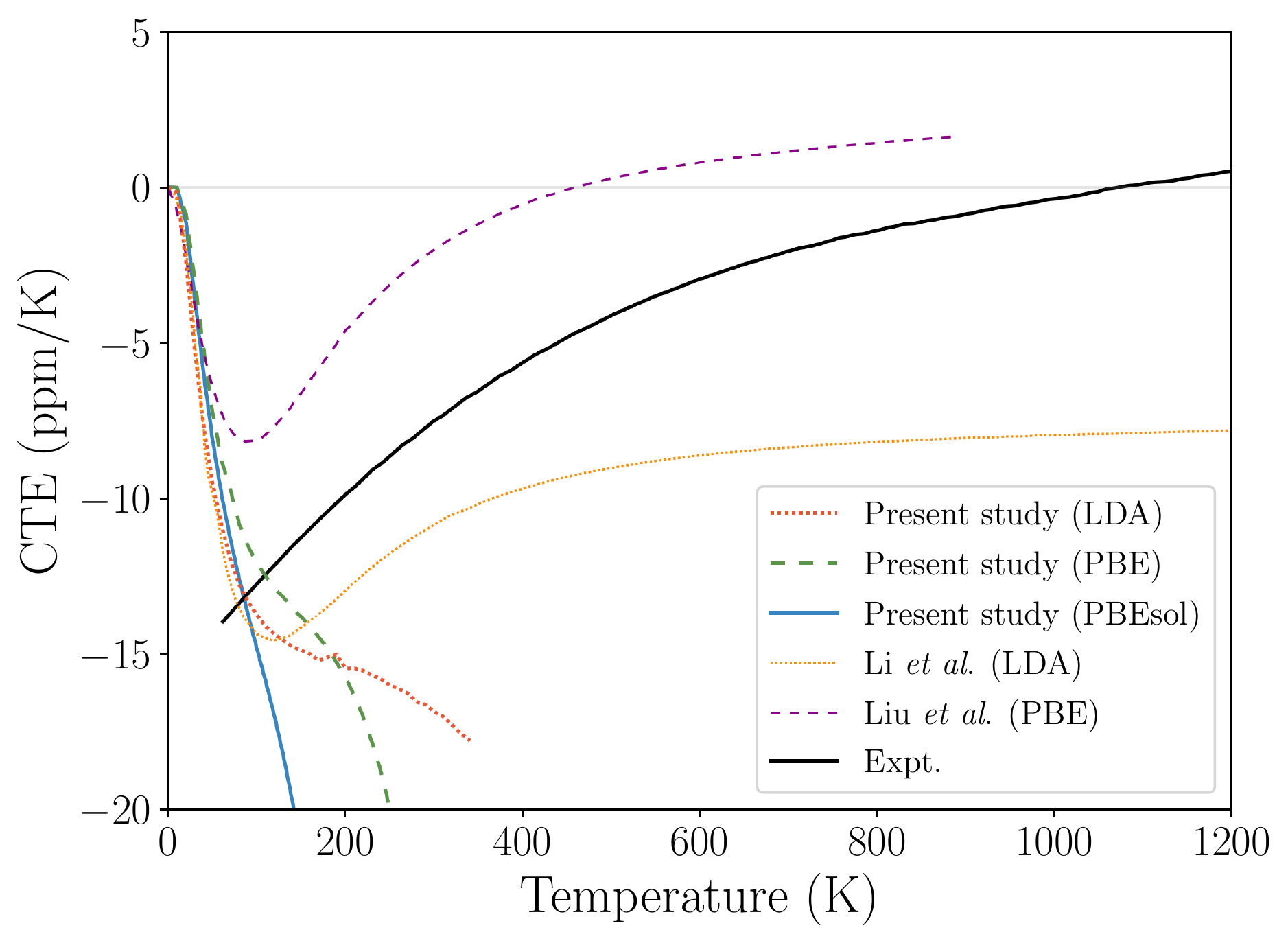}
 \caption{CTE calculated within the QHA compared with the experimental results~\cite{greve2010pronounced} and the previous QHA results~\cite{Li_Structural_2011,Liu:2015gj}. }
 \label{fig:CTE_QHA}
\end{figure}

To estimate the CTE quantitatively, we need to calculate Eq.~(\ref{eq:CTE_gamma}) or (\ref{eq:free_QHA}) from phonon frequencies. 
When an unstable mode exists, however, these equations must not be used.
One may naively expect that using $F_{\mathrm{vib}}^{(\rm{QH})}(V,T)$ only with stable phonon modes gives a reasonable estimate.
However, such a simple treatment produces an irregular oscillation of $F_{\mathrm{vib}}^{(\rm{QH})}(V,T)$ curve which affects the accuracy of the EOS fitting. The same discussion has also been made by Lan \textit{et al.}~\cite{Lan:2016gs}. 
Therefore, within the QH theory, we argue that the EOS fitting should be performed only with the volumes where all phonon modes are stable.
By this principle, we calculated the CTE with the three functionals. 
As shown in Fig.~\ref{fig:CTE_QHA}, the CTE as well as the lattice constant keep decreasing with increasing the temperature. 
Above $\sim$ 300 K, the EOS fitting failed because of the lack of valid free-energy data points in the small volume region.
In Fig.~\ref{fig:CTE_QHA}, we also compare our results with the previous QHA results~\cite{Li_Structural_2011,Liu:2015gj}, 
showing noticeable disagreement. While we cannot identify the origin of the disagreement, it can likely be attributed to the difference in the treatment of unstable phonon modes and the details of DFT calculation. Indeed, when we \tadano{performed the EOS fitting with including the incorrect $F_{\mathrm{vib}}^{(\rm{QH})}(V,T)$ data in the small volume region} \del{used $F_{\mathrm{vib}}^{(\rm{QH})}(V,T)$ calculated from stable phonon modes only}, we obtained the temperature dependence of CTE similar to the result of Ref.~\onlinecite{Liu:2015gj}.

%
\subsection{Inclusion of anharmonic free energy}\label{sec:SCPandISC}
%
The SCP theory can circumvent the limitation of the QHA mentioned above because it can stabilize the unstable soft mode by the Debye-Waller type renormalization. Since the SCP assumes the existence of well-defined phonon modes $(\Omega_{q}^{2} > 0)$, the SCP equation (\ref{eq:SCP}), if converged, always gives stable phonons. Therefore, $F_{\mathrm{vib}}^{(\mathrm{SCP})}(V,T)$ can be evaluated even when an unstable mode exists within the HA. Besides, it is straightforward to include the additional correction term from the cubic anharmonicity by evaluating Eq.~(\ref{eq:3rd}) on top of the SCP solution.

\begin{figure}[tb]
 \centering
 \includegraphics[width=8.5cm,clip]{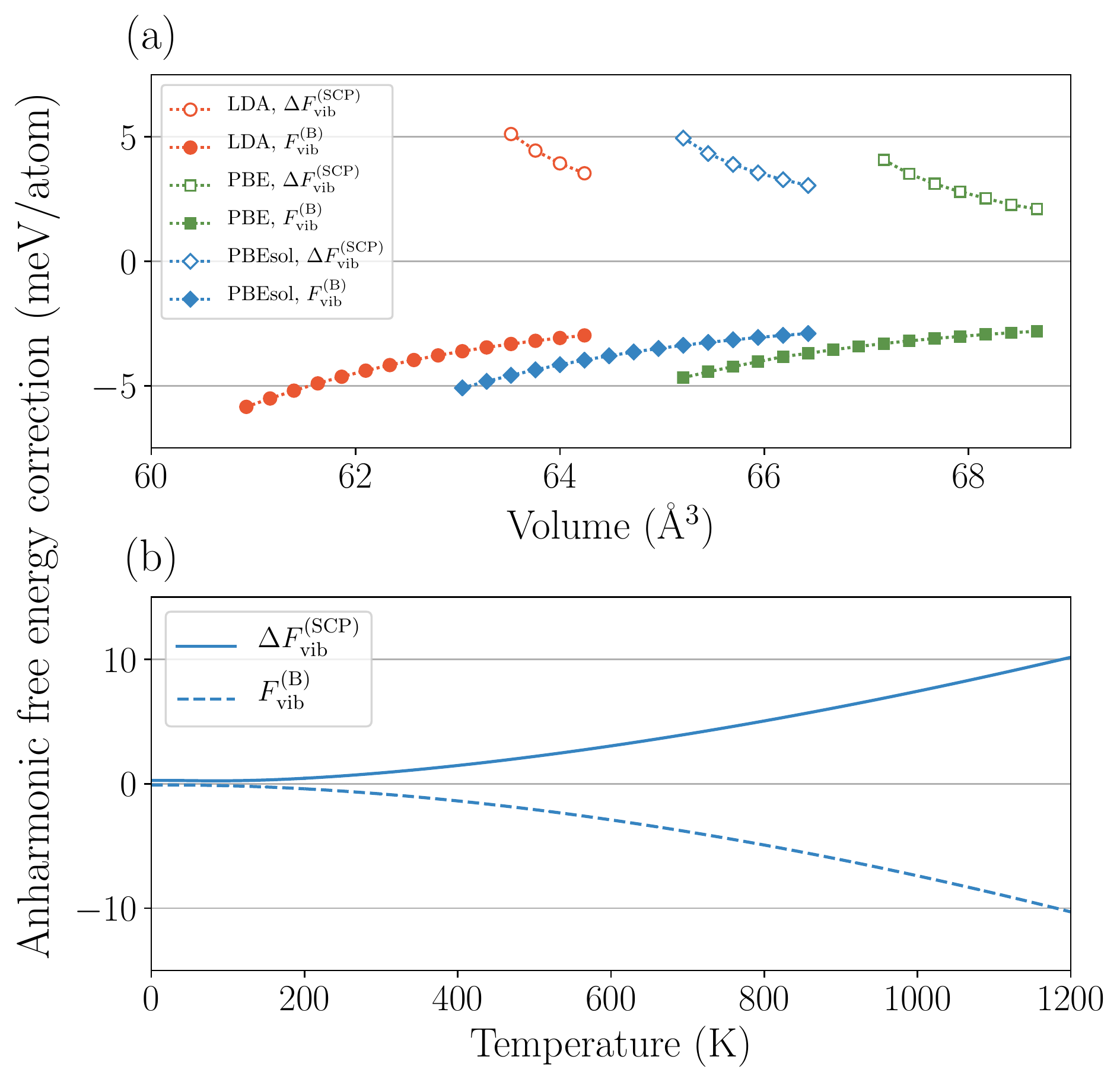}
 \caption{Anharmonic free-energy correction $\Delta F_{\mathrm{vib}}^{(\mathrm{SCP})} (V,T)$ [Eq.~(\ref{eq:Delta_FE_SCP})] and $F_{\mathrm{vib}}^{(\mathrm{B})}(V,T)$ [Eq.~(\ref{eq:3rd})]; (a) volume dependence at 600 K and (b) temperature dependence for PBEsol with $a=4.025$ \AA.}
 \label{fig:FE_AH}
\end{figure}

To quantify the effect of quartic and cubic anharmonicities, we first investigate the volume and temperature dependence of the anharmonic free energies as shown in Fig.~\ref{fig:FE_AH}. 
The anharmonic free energy correction based on the SCP is defined as
\begin{equation}
\Delta F_{\mathrm{vib}}^{(\mathrm{SCP})}(V,T) = F_{\mathrm{vib}}^{(\mathrm{SCP})}(V,T) - F_{\mathrm{vib}}^{(\mathrm{QHA})}(V,T), \label{eq:Delta_FE_SCP}
\end{equation}
which can be defined only when all phonon modes are stable within the HA.
As shown in Fig.~\ref{fig:FE_AH}(a), the $\Delta F_{\mathrm{vib}}^{(\mathrm{SCP})}(V,T)$ value of ScF$_{3}$ is positive and decreases gradually as the cell volume increases. Therefore, a larger unit-cell volume system is relatively more stabilized by the quartic anharmonicity. On the other hand, the correction from the bubble diagram $F_{\mathrm{vib}}^{(\mathrm{B})}(V,T)$ is negative and shows an opposite volume dependence. The sign of the total correction $\Delta F_{\mathrm{vib}}^{(\mathrm{SCP})}(V,T) + F_{\mathrm{vib}}^{(\mathrm{B})}(V,T)$ is positive in most of the studied volumes, but it becomes negative in the large volume systems as noticeable in the PBE results. Our result thus shows the importance of the anharmonic correction not only from the quartic anharmonicity but also from the cubic one, which compete with each other.

\begin{figure*}[tb]
 \centering
 \includegraphics[width=0.95\textwidth,clip]{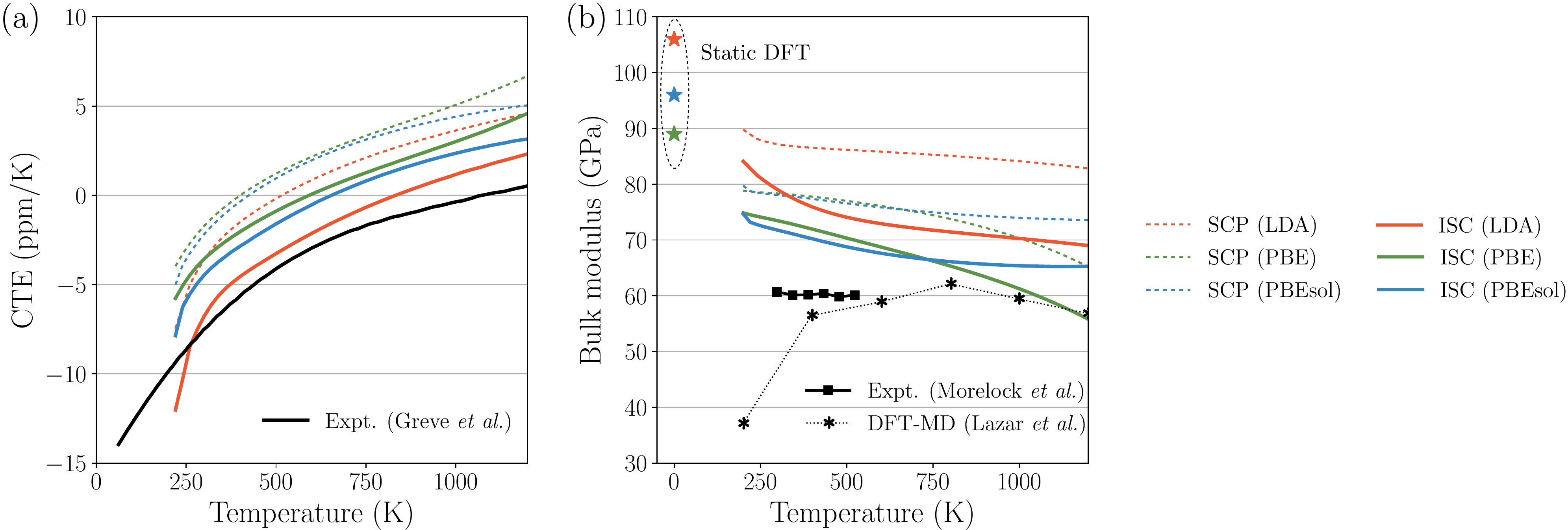}
 \caption{(a) CTE and (b) isothermal bulk modulus calculated within the SCP and ISC theories compared with the experimental results of Greve \textit{et al.}~\cite{greve2010pronounced} and Morelock \textit{et al.}~\cite{Morelock:2013gi}. \tadano{The DFT-MD result of Lazar \textit{et al.}~\cite{lazar2015negative} obtained within PBE is also shown for comparison.}}
 \label{fig:CTE_AH}
\end{figure*}

Figure \ref{fig:CTE_AH} shows the temperature-dependent CTE and bulk modulus calculated within the SCP and ISC theories.
The SCP theory correctly reproduced the positive sign of the CTE observed in the high-temperature region. However, the temperature at which the sign change occurs was around 400--500 K, which underestimates the experimental value of $\sim$1100 K.
This underestimation was partially cured by the correction from the bubble diagram, and the sign change temperature became $\sim$600-800 K.
Considering the fact that the CTE is sensitive to the exchange-correlation functional and pseudopotentials, the agreement between our ISC results with the experimental CTE of Greve \textit{et al.}~\cite{greve2010pronounced} is reasonable. 
\tadano{We expect that adopting a hybrid exchange-correlation functional and/or treating the effect of the bubble diagram fully self-consistently would improve the prediction accuracy, which are left for future study.}

\begin{figure}[tb]
  \centering
  \includegraphics[width=7.5cm,clip]{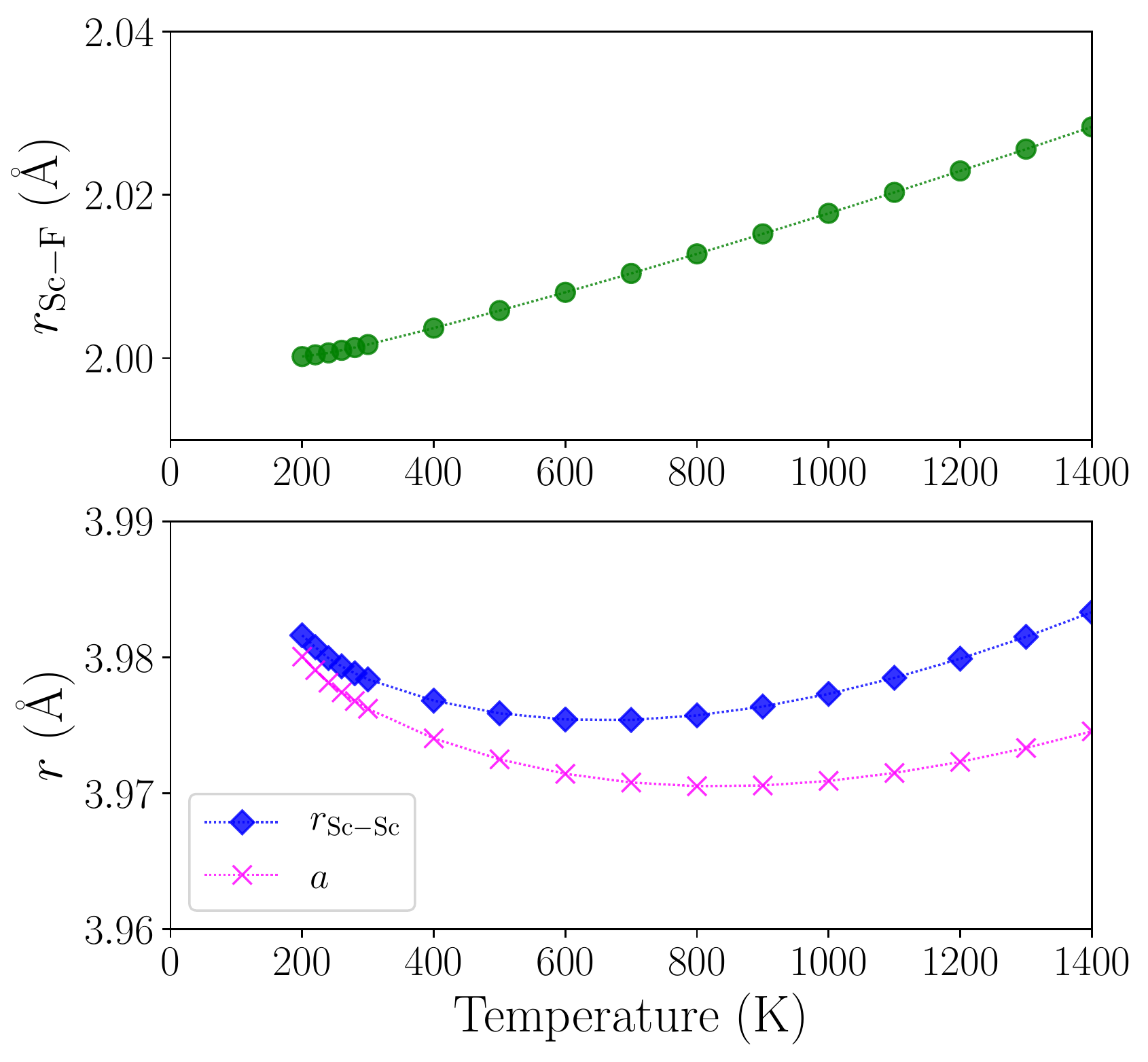}
  \caption{Upper panel: Temperature dependence of the average bond length $r$ between the nearest Sc-F pair. Lower panel: Temperature dependence of the lattice constant $a$ and the average distance $r$ between the nearest Sc-Sc pair. All values were calculated within LDA. The temperature dependence of the $a$ value was obtained from the ISC free energy, and the perpendicular MSRD at each temperature was calculated from the SCP lattice dynamics wavefunction.}
  \label{fig:average_bond_length}
\end{figure}

\ooba{(2-2) Moreover, we compared the mean square relative displacement (MSRD) of Sc-Sc and Sc-F bonds to experiment~\cite{Hu_New_2016}. The result shows that our calculation is reasonably the same as the experiment but underestimate $\gamma$ of Sc-F bond and overestimate that of Sc-Sc one. This may result from our underestimation of negative-to-positive transition temperature of CTE.}
\tadano{
  We also calculated the mean square relative displacement (MSRD) of the nearest Sc-Sc and Sc-F pairs from the SCP lattice dynamics wavefunction. Let $\Delta \bm{u} = \bm{u}_{i}-\bm{u}_{j}$ denote the difference of the instantaneous thermal displacements of two involving atoms; the parallel (perpendicular) MSRD is defined as MSRD$_{\| (\perp)} = \braket{\Delta u_{\|(\perp)}^{2}}$, where $\Delta u_{\|(\perp)}$ is the projection of $\Delta \bm{u}$ along the axis parallel (perpendicular) to the bond direction. The calculated MSRD$_{\|}$ values for the Sc-Sc and Sc-F were rather similar, whereas the MSRD$_{\perp}$ value of the Sc-F pair was much larger than that of the Sc-Sc pair. The anisotropy of the MSRD defined as $\gamma = \mathrm{MSRD}_{\perp}/\mathrm{MSRD}_{\|}$ was quite large for the Sc-F pair and amounted to $\gamma\sim 16$ in the high-temperature range. All of these results are in reasonable agreement with the experimental data of Hu \textit{et al.}~\cite{Hu_New_2016}, where the MSRD anisotropy of the Sc-F bond is reported as $\gamma \sim 19$. Moreover, we estimated the average bond length $r$ from the perpendicular MSRD as~\cite{PhysRevB.70.174301}
\begin{equation}
  r = R_{0} + \frac{\braket{\Delta u_{\perp}^{2}}}{2R_{0}},
  \label{eq:average_distance}
\end{equation}
where $R_{0}$ is the bond length at absolute rest. For the nearest Sc-Sc pair, the $r$ value shows the temperature dependence similar to that of the lattice constant $a$ since the second term of Eq.~(\ref{eq:average_distance}) is much smaller than the first term, as shown in Fig.~\ref{fig:average_bond_length} (lower panel). By constrast, the $r$ value of the nearest Sc-F increases monotonically with heating owing to the large perpendicular MSRD factor in the second term (Fig.~\ref{fig:average_bond_length}, upper panel). These trends are in accord with the findings of the previous MD and experimental studies~\cite{lazar2015negative,Hu_New_2016,piskunov2016interpretation}.}

\ooba{In addition, we semiquantitatively reproduced the weak temperature dependence of the isothermal bulk modulus by considering anharmonicity in comparison with the study of Morelock \textit{et al.}~\cite{Morelock:2013gi} as shown in Fig.~\ref{fig:CTE_AH}(b). Its overestimation is thought to be dependent on the vicinity of the structural phase transition between cubic and rhombohedral one.~\cite{lazar2015negative}}
\tadano{
The bulk modulus calculated from the Helmholtz free energy is smaller than that of the static DFT result and decreases weakly with increasing the temperature as shown in Fig.~\ref{fig:CTE_AH}(b). Also, we see that the inclusion of the cubic anharmonicity by the ISC theory reduces the bulk modulus and improves the agreement with the experimental data~\cite{Morelock:2013gi}. In the high temperature region above 750 K, our ISC result within PBE agrees well with the previous MD result based on PBE~\cite{lazar2015negative}. However, in the low temperature region, our prediction tends to overestimate the experimental bulk modulus by $\sim$10 GPa. The disagreement between our results and the previous experimental and MD studies in the low-temperature region can likely be attributed to the thermally induced distortions of the cubic phase in the vicinity of the cubic-to-rhombohedral phase transition~\cite{lazar2015negative}.
}

\ooba{(1-2) In our study, we only consider the quartic and cubic anharmonicity which is suggested to be the main contribution to the CTE of ScF3. Then the CTE derived by \textit{ab initio} molecular dynamics (AIMD) represents the experimental result more than ours.~\cite{lazar2015negative} However, we can improve the accuracy of the approximation by considering the effect of cubic anharmonicity nonperturbatively like the quartic one in Eq. (\ref{eq:F1}).}

\subsection{Quantum effect on CTE}

\begin{figure}[tb]
  \centering
  \includegraphics[width=7.5cm,clip]{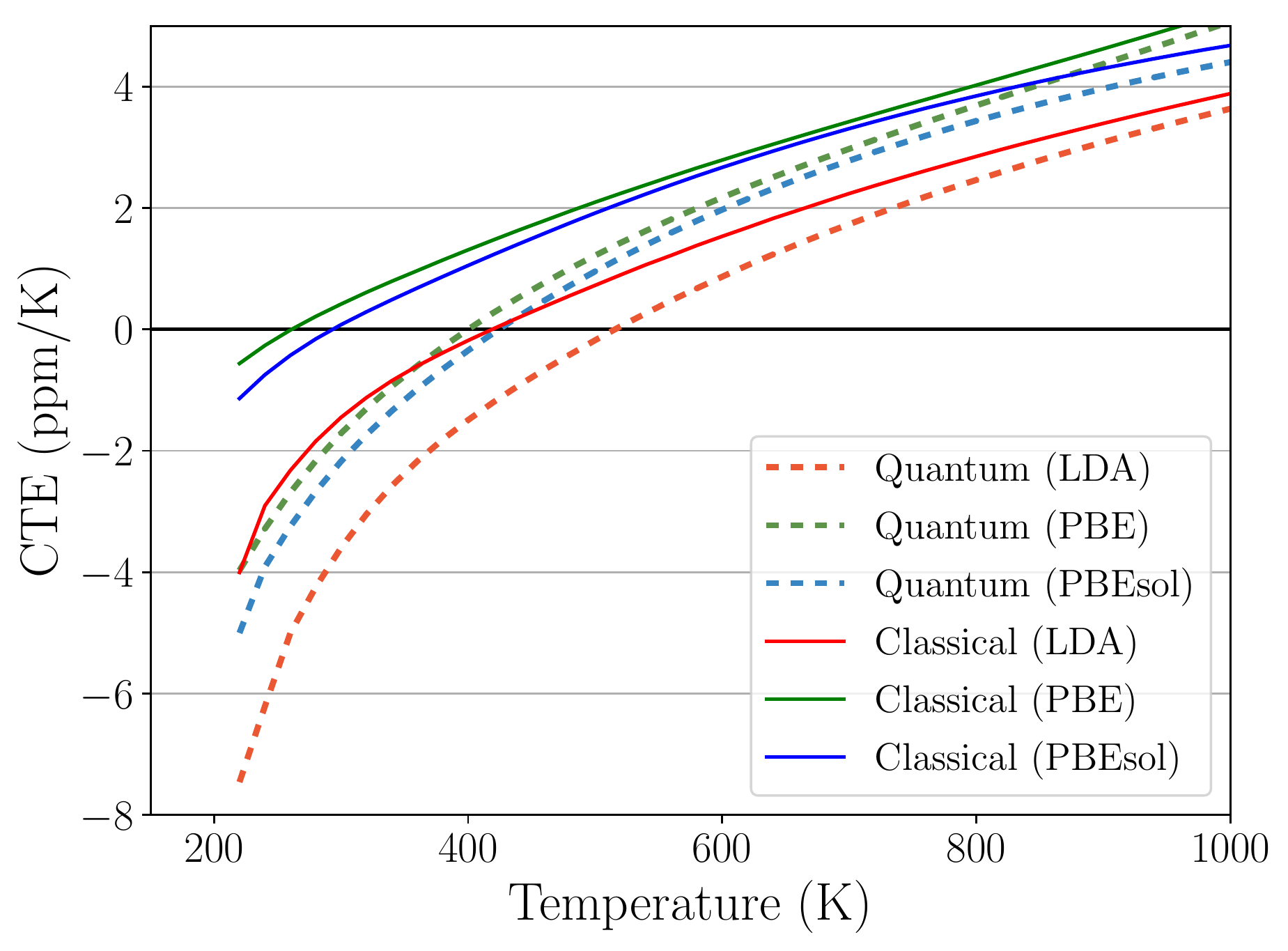}
  \caption{Comparison of the thermal expansion coefficients calculated by using the quantum and classical SCP theories. The quantum SCP theory [Eq.~(\ref{eq:Free_SCP})] correctly accounts for the nuclear quantum effects, whereas the classical SCP theory [Eq.~(\ref{eq:Free_SCP_CL})] neglects it.}
  \label{fig:CTE_classical}
\end{figure}

\ooba{(2-7) This time we investigate the quantum effect on NTE of ScF$_{3}$. For this treatment, we used the SCP approximation and calculated the CTEs by replacing the Bose-Einstein distribution function with $1/\beta\hbar\omega$ with LDA, PBE and PBEsol exchange correlational functional. We found that the CTEs without quantum effect underestimate the negative-to-positive transition temperature about 100K. It is suggested that the quantum effect is important for the strong negative thermal expansion at lower temperatures.} 

\tadano{
  Next, we discuss the nuclear quantum effects on the CTE of ScF$_{3}$. Our theoretical approaches described in Sec.~\ref{sec:theory} correctly account for the effects of the zero point motion, which are neglected in classical AIMD simulations. Since AIMD is one of the most powerful tools to study thermal expansion of materials, it should be meaningful to understand how the quantum effect can affect the CTE quantitatively.
}

\tadano{
  To quantify the nuclear quantum effects based on the SCP and ISC theories, we derived the formulas of the Helmholtz free energy in the classical limit (see Appendix \ref{sec:classical}) and employed them to evaluate the CTE of ScF$_{3}$ within the classical approximation. We then found that the classical theory systematically underestimated the magnitude of the NTE in the entire temperature range, which was particularly noticeable below $\sim$ 500 K as shown in Fig.~\ref{fig:CTE_classical}. Consequently, the transition temperature from negative to positive CTE was also underestimated by $\sim$ 100 K irrespective of the employed exchange correlation potential. Above $\sim$ 800 K, the results based on the quantum and classical statistics are almost the same. These results clearly evidence the non-negligible contribution of the nuclear quantum effect to the CTE of ScF$_{3}$ below $\sim$ 500K, where the usage of the classical AIMD is not justfied.
}

\subsection{Temperature dependence of soft mode frequency} 
\label{sec:phonons}
\begin{figure}[bt]
 \centering
 \includegraphics[width=8.5cm, clip]{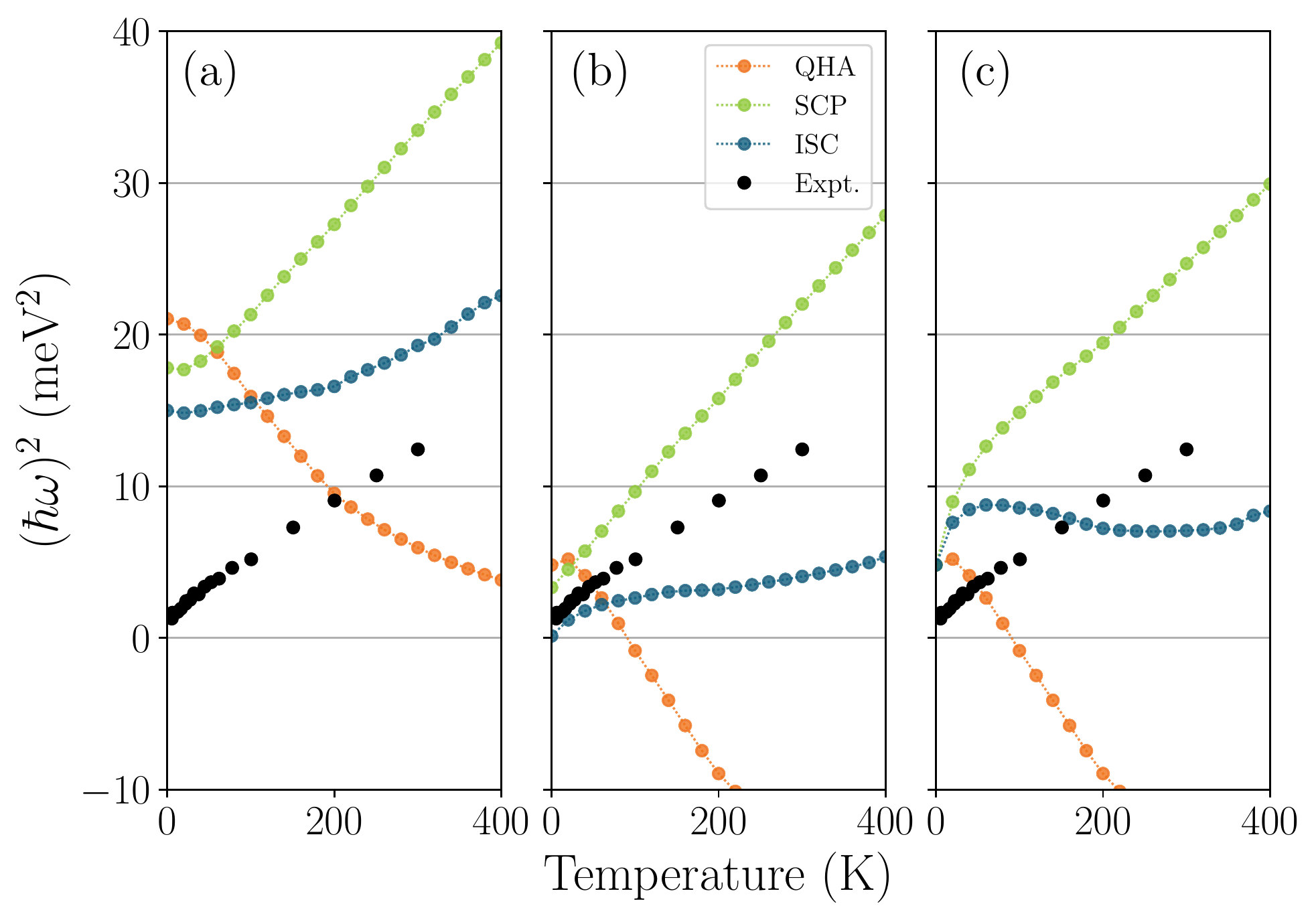} 
 \caption{Temperature dependence of squared phonon frequency of the R$^{4+}$ soft mode calculated within PBEsol in comparison with the experimental result of Handunkanda \textit{et al.}~\cite{Handunkanda:2015dc}. (a) The calculation at each temperature is conducted with the corresponding ISC volume. (b) The calculation is conducted with the lattice constant adjusted to match the experimental lattice constant at 0 K. (c) Same as (b) but the classical approximation is made.}
 \label{fig:SoftMode}
\end{figure}

%
Finally, we discuss the influence of quartic and cubic anharmonicities on the frequency of the R$^{4+}$ soft mode.
Figure \ref{fig:SoftMode} shows the temperature dependence of the squared phonon frequency of the R$^{4+}$ soft mode calculated with the PBEsol functional. The results shown in Fig.~\ref{fig:SoftMode}(a) were calculated with the equilibrium lattice constant of the ISC calculation, and the ISC phonon frequency was calculated by $\Omega_{q}^{(\mathrm{ISC})} = (\Omega_{q}^{2} + 2\Omega_{q}\Delta_{q}^{(\mathrm{B})}(\Omega_{q}))^{\frac{1}{2}}$ where $\Delta_{q}^{(\mathrm{B})}(\omega)=-\mathrm{Re}\Sigma_{q}^{(\mathrm{B})}(\omega)$. The harmonic and anharmonic IFCs at these new volumes were obtained by linearly interpolating the values calculated at the nearest two volumes. As shown in the figure, the QHA frequency softens as a function of temperature because of the NTE and the negative Gr\"{u}neisen parameter. This temperature dependence is opposite to the IXS results~\cite{Handunkanda:2015dc,Occhialini:2017cv}. The hardening of the soft mode frequency can be reproduced only by the SCP and ISC theories. Unfortunately, however, the SCP and ISC phonon frequencies systematically overestimate the experimental values.
Since the frequency is sensitive to the adopted volumes, the deviation may be mitigated by improving the agreement of the PBEsol and experimental lattice constants. To examine this point, we adjusted the external pressure $P$ in such a way that the PBEsol lattice constant agrees with the experimental result at 0 K, resulting in $P\approx 0.7$ GPa. Then, the agreement was improved considerably as shown in Fig.~\ref{fig:SoftMode}(b). In the low-temperature limit, the ISC result best agrees with the IXS results, while the SCP theory well explains the slope of the hardening. The IXS data lies in between the SCP and ISC results in Fig.~\ref{fig:SoftMode}(b). This may indicate the necessity of fully self-consistent treatment of the cubic and quartic anharmonicities, which is left for a future study.

\begin{figure}[bt]
 \centering
 \includegraphics[width=8.5cm, clip]{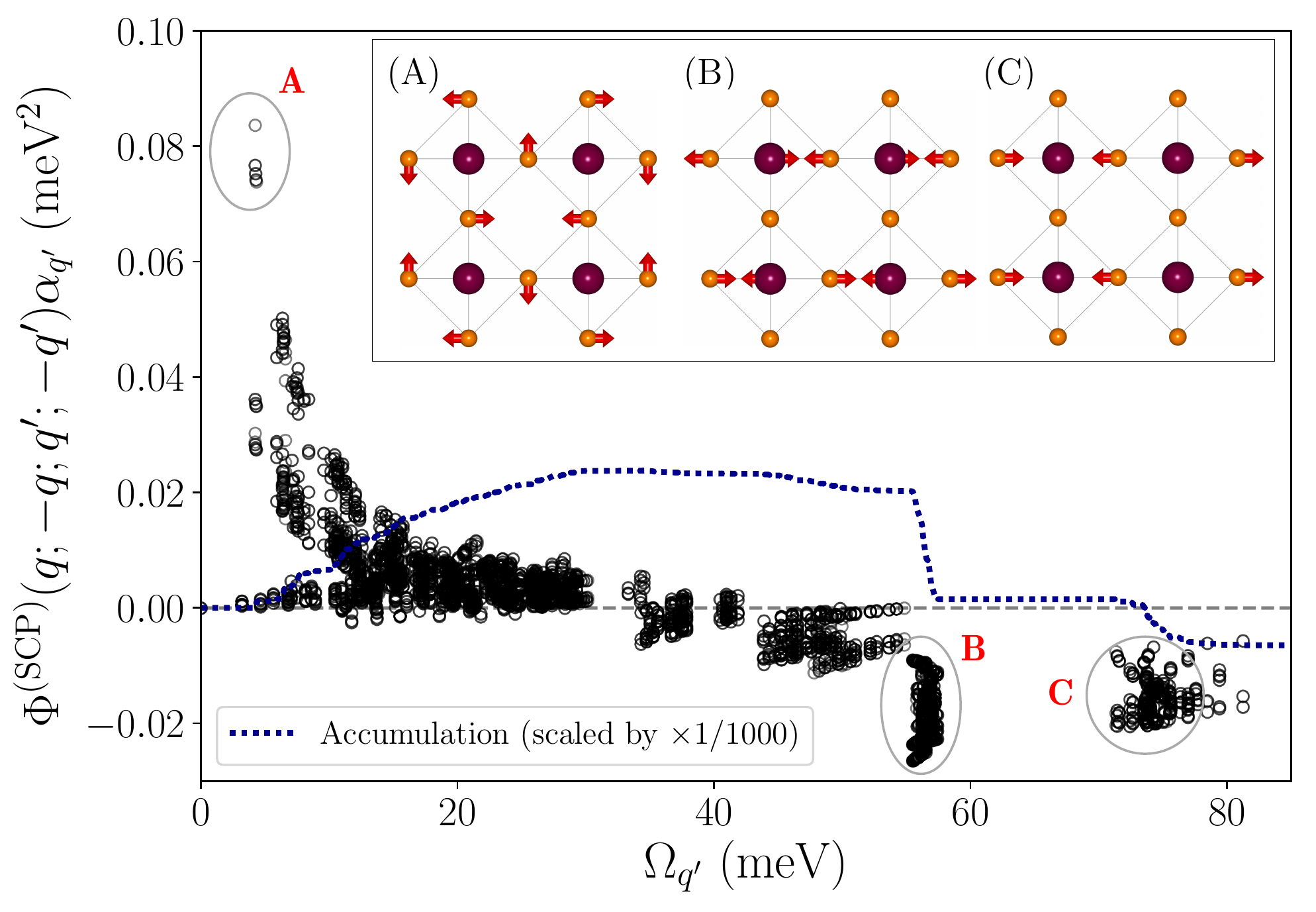} 
 \caption{Mode-dependent contribution to the quartic renormalization of the R$^{4+}$ soft mode at 0 K. The open circle shows the contribution from phonon mode $q'$ and the dotted line shows its cumulative value. The inset shows the displacement patterns of the phonon modes having large positive and negative quartic interaction with the R$^{4+}$ soft mode.}
 \label{fig:Modal_R}
\end{figure}

It is interesting to observe in Figs.~\ref{fig:SoftMode}(a) and \ref{fig:SoftMode}(b) that the SCP frequency is smaller than the QHA result near $\sim$ 0 K, meaning that the quartic anharmonicity decreases the soft mode frequency.
This behavior seems to contradict with the large positive quartic potential of the R$^{4+}$ mode~\cite{Li_Structural_2011}, which must harden the frequency. To understand this unusual behavior, we investigated the mode-dependent contribution to the quartic renormalization $\Phi^{(\mathrm{SCP})}(q;-q;q';-q')\alpha_{q'}$ [see Eq.~(\ref{eq:SCP})] for $q=R^{4+}$ and $q'$ on the 8$\times$8$\times$8 uniform grid. Figure ~\ref{fig:Modal_R} shows the mode dependence of $\Phi^{(\mathrm{SCP})}(q;-q;q';-q')\alpha_{q'}$ at 0 K. The quartic coupling is strongly positive in the low-frequency phonon modes, which are the rigid unit motion of the fluorine octahedra, in accord with the previous numerical evaluation~\cite{Li_Structural_2011}. In contrast, the coupling coefficients are negative for high-frequency optical modes in 40--80 meV, which are mostly dominated by the vibration of fluorine atoms as shown in the inset of the figure. Since the total negative contribution is slightly larger than the total positive one, the SCP frequency becomes smaller than that of the QHA at 0 K. With increasing the temperature, the factor $\alpha_{q'}$ increase more rapidly for low-frequency phonon mode, and the total contribution becomes positive around 60 K [see Fig.~\ref{fig:SoftMode}(a)]. We note that this unique phenomenon results from the zero-point motion of atoms. If we replace $n_{q}$ with $n_{q} - \frac{1}{2}$ in Eqs.~(\ref{eq:SCP}) and (\ref{eq:Self_2nd}), the SCP frequency is larger than the QHA frequency irrespective of the temperature as shown in Fig.~\ref{fig:SoftMode}(c).

\section{CONCLUSION}
\label{sec:conclusion}

We investigated the role of the cubic and quartic anharmonicities in the large NTE of cubic ScF$_{3}$ by using many-body theoretical approaches based on first-principles anharmonic force constants. 
We showed that the quartic anharmonicity, which was included by the SCP theory, is essential to reproduce the observed transition from negative to positive CTE as heating. In addition, we newly found that the inclusion of the cubic anharmonicity, which was achieved by the ISC theory, improves the quantitative agreement between theoretical and experimental thermal expansivities. Therefore, both cubic and quartic anharmonicities are equally important for a quantitative understanding of the unusually large NTE of ScF$_{3}$. 
Moreover, we calculated the temperature dependence of the R$^{4+}$ soft mode within the SCP and ISC theories and obtained results that agree semiquantitatively with the IXS data, particularly when the experimental lattice constant is employed.
We showed that the R$^{4+}$ mode softens by the quartic anharmonicity near 0 K, and revealed that it results from the zero-point vibration of atoms and the negative quartic coupling between the R$^{4+}$ mode and high-frequency optical modes.

The present first-principles approach enables us to obtain the vibrational free-energy of solids with \tadano{the effects of the nuclear zero-point motion,} cubic and quartic anharmonicities, and it is even applicable to systems where unstable phonon modes exist within the HA (e.g., high-temperature phase of solids). Since the method is based on the reciprocal space formalism, thermodynamic properties in the thermodynamic limit can be calculated efficiently by using the interpolation technique, which is particularly important for a computational estimation of CTE and phase boundaries.

\begin{acknowledgments}
We would like to acknowledge the support by the MEXT Element Strategy Initiative to Form Core Research Center in Japan.
T.T. is partly supported by JSPS KAKENHI Grant No. 16K17724 and ``Materials research by Information Integration'' Initiative (MI2I) project
of the Support Program for Starting Up Innovation Hub from Japan Science and Technology Agency (JST).
The computation in this work has been done using the facilities of the Supercomputer Center, Institute for Solid State Physics, The University of Tokyo.
\end{acknowledgments}

\appendix
\section{TAYLOR SERIES EXPANSION OF POTENTIAL ENERGY}

\label{sec:TEP}

If the atomic displacements are small compared with the interatomic distance,
the potential energy of the interacting atomic system can be expanded in a power series 
of the displacements $\bm{u}(st) = \bm{R}(st) - \bm{R}^{0}(st)$ as 
\begin{equation} 
U = U_0 + U_2 + U_3 + U_4 + \cdots, \label{eq:U}
\end{equation}
where
\begin{align}
U_n &= \frac{1}{n!}\sum_{\{s, t, \mu\}} \Phi_{\mu_1 \cdots \mu_n}(s_1 t_1;\cdots; s_n t_n) \notag \\
& \hspace{20mm} \times u_{\mu_1}(s_1 t_1)\cdot\cdot\cdot u_{\mu_n}(s_n t_n).
\label{eq:Un}
\end{align}
Here, $\mu=x,y,z$ and $u_{\mu}(st)$ is the displacement of atom $s$ in the $t$th cell.
The coefficient $\Phi_{\mu_1 \cdot\cdot\cdot \mu_n}(s_1 t_1;\cdot\cdot\cdot; s_n t_n)$ is the
$n$th-order derivative of $U$ with respect to atomic coordinates, 
which is called $n$th-order interatomic force constant (IFC).
Next, we introduce the complex normal coordinate $Q_{q}$, with which the atomic displacement is expressed as
\begin{equation} 
u_\mu(st)=(NM_{s})^{-\frac{1}{2}}\sum_{q}Q_{q}e_\mu (s, q) e^{i\bm{q}\cdot\bm{r}(t)}. \label{eq:u_in_Q}
\end{equation}
By substituting Eq.~(\ref{eq:u_in_Q}) for Eq.~(\ref{eq:Un}), we obtain $U_{n}$ expressed in terms of the normal coordinate as follows:
\begin{align}
U_n &= \frac{1}{n!}\sum_{\{q\}} \Delta(\bm{q}_{1}+\cdots+\bm{q}_{n})\Phi(q_{1};\dots;q_{n}) Q_{q_{1}}\cdots Q_{q_{n}}, \label{eq:Un_in_Q}
\end{align}
where
\begin{align} \label{eq:Fourier}
&\Phi(q_{1};\dots;q_{n}) = N^{1-\frac{n}{2}} \sum_{\{s, \mu\}}\frac{e_{\mu_1}(s_1, q_1)\cdots e_{\mu_n}(s_n, q_n)}{\sqrt{M_{s_1} \cdots M_{s_n}}} \notag \\
&\times \sum_{t_2 \cdots t_n} \Phi_{\mu_1 \cdots \mu_n}(s_{1}0;\cdots; s_{n}t_{n}) e^{i (\bm{q}_2 \cdot \bm{r}(t_2)+\cdots+\bm{q}_n \cdot \bm{r}(t_n))}.
\end{align}
When phonon frequencies of all phonon modes are real in the entire Brillouin zone, one may further transform Eq.~(\ref{eq:Un_in_Q}) into a second quantization representation by using $Q_{q} = (\hbar/2\omega_{q})^{1/2}A_{q}$ with $A_{q} = b_{q} + b_{-q}^{\dagger}$ being the displacement operator.

In Eqs.~(\ref{eq:u_in_Q})--(\ref{eq:Fourier}), we have introduced $U_{n}$ expressed in terms of the normal coordinate $Q_{q}$ within the HA.
Instead of using the harmonic eigenvectors, one can also use the SCP eigenvectors and associated normal coordinates $\tilde{Q}_{q}$ for $U_{n}$ as
\begin{align}
U_n &= \frac{1}{n!}\sum_{\{q\}} \Delta(\bm{q}_{1}+\cdots+\bm{q}_{n})\Phi^{(\mathrm{SCP})}(q_{1};\dots;q_{n}) \tilde{Q}_{q_{1}}\cdots \tilde{Q}_{q_{n}}. \label{eq:Un_in_Q2}
\end{align}
Since the SCP eigenvector is a unitary transformation of the harmonic one, 
i.e. $\varepsilon_{\mu}(s,\bm{q}j)=\sum_{k}e_{\mu}(s,\bm{q}k) (C_{\bm{q}})_{kj}$, it is easy to show that the following transformation rule holds:
\begin{align}
\Phi^{(\mathrm{SCP})}(\bm{q}_{1}j_{1};\dots;\bm{q}_{n}j_{n}) &= \sum_{\{k\}} (C_{\bm{q}})_{k_{1}j_{1}}\cdots (C_{\bm{q}})_{k_{n}j_{n}} \notag \\
& \times \Phi(\bm{q}_{1}k_{1};\dots;\bm{q}_{n}k_{n}).
\end{align}

\tadano{
\section{SCP THEORY AND VIBRATIONAL FREE ENERGY IN THE CLASSICAL LIMIT}
\label{sec:classical}
In the classical limit ($\hbar\rightarrow 0$), the Bose--Einstein distribution function $n_{\mathrm{B}}(\omega) = 1/(e^{\beta\hbar\omega}-1)$ can be replaced with $n_{\mathrm{C}}(\omega) = 1/\beta\hbar\omega$, and the constant terms appearing with $n_{\mathrm{C}}(\omega)$ can be omitted because they are negligible compared with $n_{\mathrm{C}}(\omega)$. By this systematic modification, the free-energy expressions in the classical limit can be obtained straightforwardly. 
The QH free energy is given as 
\begin{equation} 
  F_{\mathrm{vib}}^{(\rm{QH},\rm{CL})}(V,T) = \frac{1}{\beta} \sum_{q} \ln{(\beta\hbar\omega_{q})} . \label{eq:free_QHA_CL}
\end{equation}
The original SCP equation [Eq.~(\ref{eq:SCP})] is still valid in the classical limit, but the temperature dependent factor $\alpha_{q}$ must be replaced with $\alpha_{q}^{\mathrm{C}} = \hbar n_{\mathrm{C}}(\Omega_{q})/\Omega_{q} = 1/(\beta\Omega_{q}^{2})$. Then, the SCP free energy becomes
\begin{align}
  F_{\mathrm{vib}}^{(\mathrm{SCP},\mathrm{CL})}(V,T) &= \frac{1}{\beta} \sum_{q} \ln{(\beta\hbar\Omega_{q})} \notag \\
  & - \frac{1}{4}\sum_{q} \left[ \Omega_{q}^{2} (V,T) - (C_{\bm{q}}^{\dagger}\Lambda_{\bm{q}}^{(\mathrm{HA})}C_{\bm{q}})_{jj}  \right] \alpha_{q}^{\mathrm{C}}.  \label{eq:Free_SCP_CL}
\end{align}
The ISC free energy [Eq.~(\ref{eq:FE_ISC})] has first and second order terms with respect to $n_i$ in the numerator, where $n_{i}  = n_{\mathrm{B}}(\Omega_{i})$. In the classical limit, the first order terms can be neglected and $n_{\mathrm{B}}(\Omega_{i})$ is approximated to the classical form $n'_{i} = n_{\mathrm{C}}(\Omega_{i})$. Therefore, we obtain
\begin{widetext}
  \begin{align} 
  F_{\mathrm{vib}}^{(\rm{B},\mathrm{CL})}(V,T) 
  &= -\frac{\hbar^2}{48} \sum_{q_{1},q_{2},q_{3}} \frac{\left | \Phi^{(\mathrm{SCP})} (q_{1};q_{2};q_{3}) \right |^2}{\Omega_{q_{1}}\Omega_{q_{2}}\Omega_{q_{3}}} \Delta(\bm{q}_1+\bm{q}_2+\bm{q}_3) 
  \Biggl[\frac{n'_1(n'_2+n'_3) + n'_2 n'_3}{\Omega_{q_{1}}+\Omega_{q_{2}}+\Omega_{q_{3}}} 
  + 3 \frac{n'_1(n'_2 + n'_3) - n'_2 n'_3 }{-\Omega_{q_{1}}+\Omega_{q_{2}}+\Omega_{q_{3}}}\Biggl]. 
  \label{eq:FE_B_CL}
  \end{align}
\end{widetext}
}

\begin{thebibliography}{54}
\bibitem [{PP_()}]{PP_note}%
  \BibitemOpen
  \href@noop {} {}\bibinfo {note} {We employed the \textsc{VASP} POTCAR files
  named \texttt{Sc\_sv} and \texttt{F}.}\BibitemShut {Stop}%
\end{thebibliography}

%

\end{document}